\documentclass[runningheads,10pt]{llncs}
\usepackage{todo}
\usepackage{times}
\usepackage{graphicx}
\let\subparagraph\paragraph
\usepackage[compact]{titlesec}
\titlespacing{\section}{0em}{0.5em}{0.5em}
\titlespacing{\subsection}{0pt}{0.3em}{0.3em}
\titlespacing{\paragraph}{0pt}{0pt}{0pt}
\usepackage[dvipsnames]{xcolor}
\usepackage{hyperref}
\hypersetup{
    bookmarks=true,         
    unicode=false,          
    pdftoolbar=true,        
    pdfmenubar=true,        
    pdffitwindow=false,     
    pdfstartview={FitH},    
    pdftitle={My title},    
    pdfauthor={Working Group 1},     
    pdfsubject={Activity Report},   
    pdfcreator={Working group 1},   
    pdfproducer={Producer}, 
    pdfkeywords={runtime verification, monitoring, taxonomy, instrumentation, testing, model-checking, assertion checking, learning}, 
    pdfnewwindow=true,      
    colorlinks=true,       
    linkcolor=red,          
    citecolor=ForestGreen,        
    filecolor=magenta,      
    urlcolor=cyan           
}

 \newcommand{\specialcell}[2][c]{%
  \begin{tabular}[#1]{@{}c@{}}#2\end{tabular}}
\begin{document}
%
\title{%
COST Action IC 1402 ArVI:\\Runtime Verification Beyond Monitoring}
\subtitle{Activity Report of Working Group 1\\ - \\Core Aspects of Monitoring}
\author
	{
	Wolfgang Ahrendt\inst{1}
	\and
	Cyrille Artho\inst{2}
	\and
	Christian Colombo\inst{3}
	\and
	Yli\`es Falcone\inst{4}
	\and
	Srdan~Krstic\inst{5}
	\and
	Martin Leucker\inst{6}
	\and
	Florian Lorber\inst{7}
	\and
	Joao Louren\c{c}o\inst{8}
	\and
	Leonardo Mariani\inst{9}
	\and
	C\'esar S\'anchez\inst{10}
	\and
	Gerardo Schneider\inst{11}
	\and
	Volker Stolz\inst{12}
	}
\institute
	{
	Chalmers University of Technology, Gothenburg, Sweden
	\and
	KTH Royal Institute of Technology, Stockholm, Sweden
	\and
	University of Malta, Msida, Malta
	\and
	Univ. Grenoble Alpes, CNRS, Inria, LIG, 38000 Grenoble, France
	\and
	Institute of Information Security, Department of Computer Science, ETH Z\"urich, Switzerland
	\and
	University of L\"ubeck, L\"ubeck, Germany
	\and
	Department of Computer Science, Aalborg University, Aalborg, Denmark
	\and
	NOVA LINCS, DI, FCT, NOVA University Lisbon, Lisbon, Portugal
	\and
	University of Milano - Bicocca, IT-20126 Milan, Italy
	\and
	IMDEA Software Institute, Madrid, Spain
	\and
	University of Gothenburg, Gothenburg, Sweden
	\and
	Western Norway University of Applied Science, Bergen, Norway
	}
\titlerunning{COST Action IC 1402 ArVI - Activity Report}
\authorrunning{Working Group 1 - Core Aspects of Monitoring}
%
\maketitle
%
%
\begin{abstract}
This report presents the activities of the first working group of the COST Action ArVI, Runtime Verification beyond Monitoring.
The report aims to provide an overview of some of the major core aspects involved in Runtime Verification.
Runtime Verification is the field of research dedicated to the analysis of system executions.
It is often seen as a discipline that studies how a system run satisfies or violates correctness properties.

The report exposes a taxonomy of Runtime Verification (RV) presenting the terminology involved with the main concepts of the field.
The report also develops the concept of instrumentation, the various ways to instrument systems, and the fundamental role of instrumentation in designing an RV framework.
We also discuss how RV interplays with other verification techniques such as model-checking, deductive verification, model learning, testing, and runtime assertion checking.
Finally, we propose challenges in monitoring quantitative and statistical data beyond detecting property violation.
\end{abstract}
%
%
%
%
%
%
\section{Introduction}
\label{sec:intro}
%
Runtime Verification (RV), as a field of research, is referred to by many names such as runtime monitoring, trace analysis, dynamic analysis, passive testing, runtime enforcement etc. (see~\cite{HavelundG05,LeuckerS09,FalconeHR13,BartocciFFR18,Falcone10,FalconeMRS18} for tutorials).
The term \emph{verification} implies a notion of \emph{correctness} with respect to some property. This is somewhat different from the term \emph{monitoring} (the other popular term) which only suggests that there is some form of behaviour being observed.
Some view the notion of monitoring as being more specific than that of verification as they take it to imply some interaction with the system, whereas verification is passive in nature. 

RV is a lightweight, yet rigorous, formal method that complements classical exhaustive verification techniques (such as model checking and theorem proving) with a more practical approach that analyses a single execution trace of a system.
At the expense of a limited execution coverage, RV can give very precise information on the runtime behaviour of the monitored system.
The system considered can be a software system, hardware or cyber-physical system, a sensor network, or any system in general whose dynamic behaviour can be observed. The archetypal analysis that can be performed on runtime behaviour is to check for correctness of that behaviour.
However, there are many other analyses (e.g., falsification analysis) or activities (e.g., runtime enforcement) that can be performed, as it will be discussed elsewhere in this report.
RV is now widely employed in both academia and industry both before system deployment, for testing, verification, and debugging purposes, and after deployment to ensure reliability, safety, robustness and security.

The RV field as a self-named community grew out of the RV workshop established in 2001, which became a conference in 2010 and occurs each year since then.
In 2014, we have initiated the international Competition on (Software for) Runtime Verification (CRV)~\cite{BartocciBF14,FalconeNRT15,RegerHF16,Bartocci2017} with the aim to foster the comparison and evaluation of software runtime verification tools. 
In 2016 and 2018, together with other partners of ARVI, we have also started to organize the two first of a series of Schools on RV~\cite{ColomboF16,Falcone18}.
%
%
\section{A Taxonomy of Runtime Verification}
%
\begin{figure}[t]
	\centering
	\includegraphics[width=.9\linewidth]{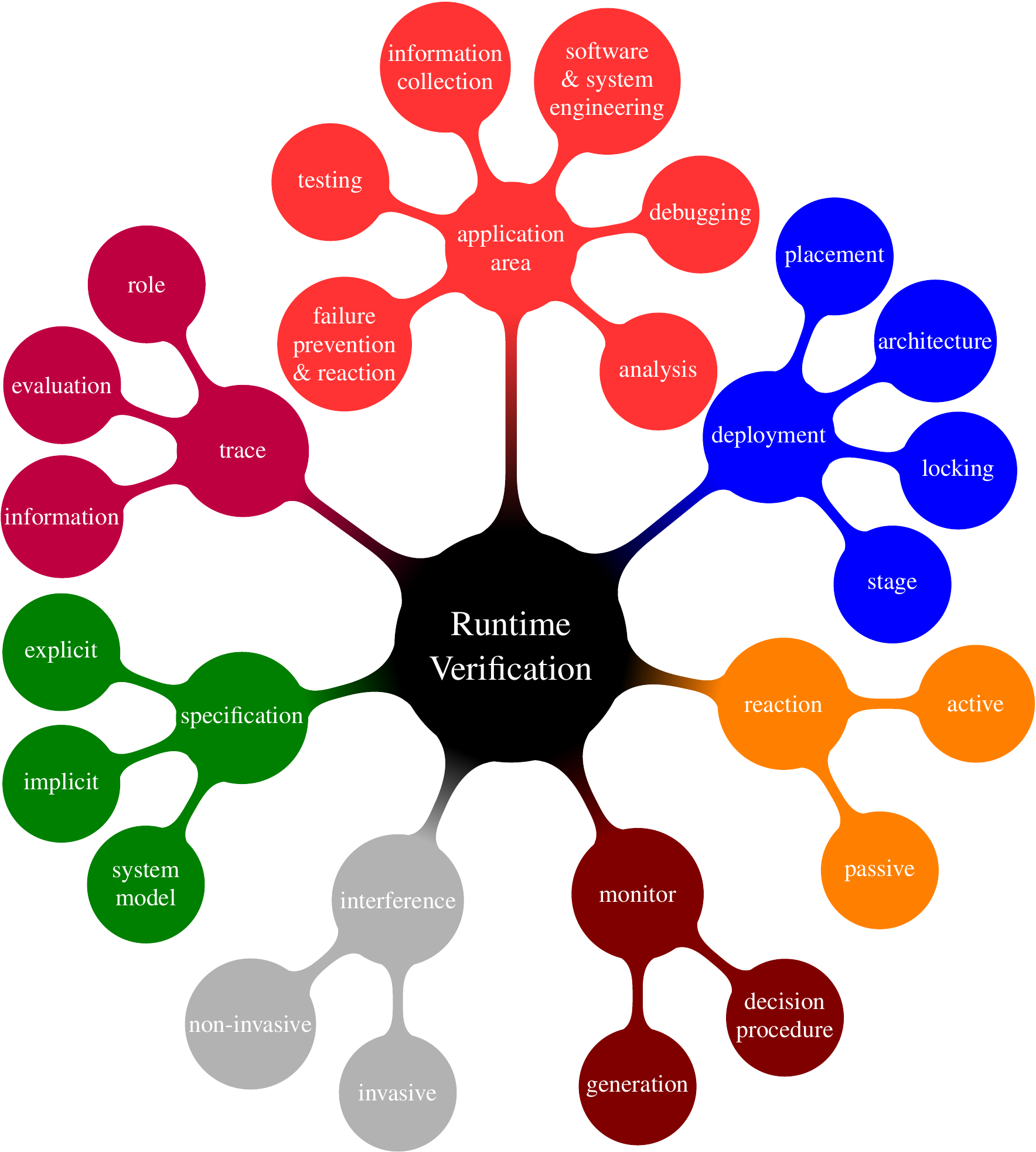}
	\caption{Mindmap overviewing the taxonomy of Runtime Verification~\cite{Taxonomy}.}
	\label{fig:taxonomy:all}
\end{figure}

Runtime Verification (RV) has grown into a diverse and active field during the last 15 years and has stimulated the development of numerous theoretical frameworks and tools.
The paper in~\cite{Taxonomy} presents a high-level taxonomy of RV concepts and use it to classify RV tools.
The classification and discussion related to RV tools is beyond the scope of this report.
We instead briefly recall the main points of the classification and refer to~\cite{Taxonomy} for further details.
We also do not discuss the application area part of the taxonomy, as applications of runtime verification is the object of study of another working group.
The taxonomy provides a hierarchical organization of the six major concepts used in the field and serves to classify several of the existing tools.
In this report, we report only on the first two levels for readability and brevity reasons:
\paragraph{Specification. }
A specification indicates the intended system behavior (property), that is \emph{what one wants to check} on the system behavior.
	It is generally one of the main inputs of a runtime verification framework designed before running the system.
	A specification exists within the context of a general system model i.e., the abstraction of the system being specified.
	A specification itself can be either implicit or explicit.
	An implicit specification is used in a runtime verification framework when there is a general understanding of the particular desired behavior.
	An explicit specification is one provided by the user of the runtime verification framework.
\paragraph{Monitor. }
	A monitor is a main component of a runtime verification framework.
	By monitor, we refer to a component executed along the system for the purposes of the runtime verification process.
	A monitor implements a semi-decision procedure which produces the expected output (either the related information for an implicit specification or the specification language output for an explicit specification).
	Note, the monitor may run forever without producing a verdict.
	Monitors must be generated from a specification. 
\paragraph{Deployment. }
	By deployment, we refer to how the monitor is effectively implemented, organized, how it retrieves the information from the system, and when it does so.

	The notion of stage describes when the monitor operates, with respect to the execution of the system.
	The notion of placement describes where the monitor operates, with respect to the running system. Therefore, this concept only applies when the stage is online. 
	The architecture of the monitor may be centralized (e.g., in one monolithic procedure) or decentralized (e.g., by utilising communicating monitors).
\paragraph{Reaction. }
	By reaction, we refer to how the monitor affects the execution of the system.
	Reaction is said to be passive when the monitor does not influence or minimally influences the initial execution of the program.
	Reaction is said to be active when the monitor affects the execution of the monitored system.
\paragraph{Trace. }
	The notion of trace appears in two places in a runtime verification framework and this distinction is captured by the role concept.
	A monitor can receive different sorts of information from a system (e.g., events, states, or signals).
	 The system evaluation of the system may refer either to specific points in time or refer to intervals of time.
\paragraph{Invasive. }
	 In absolute, a non-invasive monitoring framework being impossible, the distinction between invasive vs non-invasive better corresponds in reality to a spectrum.
	There are two sources of interference for a monitoring framework with a system: the effect of the instrumentation applied to the system and the monitor deployment.

%
%

%
%
%
%
\section{Instrumentation}
%
The term instrumentation refers to the mechanism employed to probe and to extract signals, traces of events and other information of interest from a software or hardware system during its execution.
Instrumentation determines what aspects of the system execution are made visible (to the monitor) for analysis, in terms of what elements of the computation is reported (e.g., computation steps and the data associate with them), and the relationships between the recorded events (e.g. provide a partial or total ordering, or guaranteeing that the reported event order corresponds to the order in which the respective computational step occurred).
Instrumentation also dictates how the system and the monitor execute in relation to one another in a monitoring setup.
It may either require the system to terminate executing before the monitor starts running (offline), interleave the respective executions of the system and the monitor within a common execution thread (online), or allocate the monitor and the system separate execution threads (inline vs outline); instrumentation may dictates how tightly coupled these executions need to be (synchronous vs asynchronous).  


The choice of instrumentation techniques depends on the type of system to be monitored.
For example, monitoring hardware system may require probing mixed-analog signals using physical wires, while for software the instrumentation method is strictly related to the programming language in which the software is implemented or to the low-level language in which it is compiled (i.e., bytecode, assembly, etc.).

The following instrumentation mechanisms exist:

\subsubsection{Logging or manual instrumentation}

Most systems already log important actions. Typically, the level of detail
of these logs is configurable. When a system is configured to log data
a high level of detail, the log may
contain enough information to derive a verification verdict from its
data~\cite{jepsen,naldurg2004temporal,hu2011automating}.
If this is not the case, additional data has to be
obtained by manual instrumentation (manually inserting logging code).
Furthermore, the format of the data is typically unstructured or semi-structured,
and typically have to be pre-processed before they can be used for
monitoring~\cite{zhu2010incremental,tang2011logsig}.
The advantages of this approach are that the data is typically already
available, and no special tools have to be set up.

\subsubsection{Code instrumentation (transpilation/weaving)}

With code instrumentation, the code of the program is
modified such that the necessary statements to capture
the information for runtime monitoring are inserted. This modification
is usually automatic and supported by a tool. This has the
advantage that it is possible to cover similar properties in a uniform
way across a large program, as the code is modified in a systematic
and automatic way. Furthermore, this technique is almost unlimited
w.\,r.\,t.\ the amount and type of data it can access, as code
instrumentation can be very fine-grained.

The tools that modify the code are often called transpilers (if the source
code is modified) or weavers (if source or compiled code is modified).
Transpilation occurs right before compilation, whereas compiled code
can be modified after compilation, or at load time. In all of these cases,
the interplay of various stages of building and deployment of software
may not readily accept a code instrumentation stage without adaptation,
due to actions such as code generation, dynamic loading of libraries,
other code rewriting frameworks such as Spring~\cite{johnson2004spring}, etc.
While the result of source-to-source transformed can be readily inspected,
weavers for compiled code are more flexible and also work without the source.
However, it can be difficult to ensure that all information is captured
fully and correctly, without altering the original behavior of the program.

Popular transpilers can be implemented as libraries~\cite{pawlak:hal-01169705}, term
rewriting systems \cite{balland2007tom,visser2001stratego},
compiler extensions~\cite{lee2003cetus,gcc-instr},
domain-specific languages~\cite{klint2009rascal},
or even transformation generators~\cite{kuipers2001object}.
Weavers include domain-specific languages such as AspectJ~\cite{kiczales2001overview} or AspectC++\cite{spinczyk2002aspectc++},
but also libraries such as ASM~\cite{bruneton02}.
Code instrumentation is also often provided by custom tools that have
been designed with the given verification task in
mind~\cite{rubanov2011runtime,nethercote2007valgrind,Savage:1997:EDD:269005.266641}.

\subsubsection{Call interception}

When analyzing activity such as input/output, which is accessed via
existing libraries, the easiest way to observe these actions is by
overloading the libraries with a wrapper. The wrapper then intercepts
the call and updates
the runtime monitor before or after calling the original library code.
This approach differs from code instrumentation in that code is not
modified throughout the system, but instead, the modification is
made by intercepting calls systematically, and providing additional
functionality before or after each function call.

Typical mechanisms to achieve this are the usage of
the linker to replace a library call with a wrapper~\cite{gcc-link},
kernel modifications~\cite{artho2015using},
the boot-classpath in Java (up to Java 8)~\cite{java8-bootcp}
or Java's module system~\cite{java-modules}. This approach is less
flexible than other approaches, in that it can only modify the
behavior of code at function call boundaries,
but it is easy to confine the
modifications to small parts of the system. However, not all platforms
have a straightforward mechanism of keeping the unmodified version
and delegating a call to the original code inside the wrapper;
for instance, older versions of Java completely replaced the overloaded
library without any way of accessing the old code. Furthermore, the mechanisms
to overload libraries are very specific to each platform, so tools
using this technique are not portable.

\begin{table}[t]
\caption{Advantages and disadvantages of different approaches\label{tab:instr}}
\centering
\begin{tabular}{l l}
\hline
\textbf{Approach} & \textbf{Advantages and disadvantages}\\
\hline
Logging& $+$ Much data already available; no special tooling needed.\\
& $-$ Data format is unstructured; available data may be limited; ad hoc.\\
\hline
Code instrumentation& $+$ Systematic, flexible approach.\\
& $-$ Difficult to use and test; may interfere with original program.\\
\hline
Call interception& $+$ Limited modifications; good performance.\\
&$-$ Not portable; limited to the interfaces of libraries.\\
\hline
Execution environment& $+$ Fast; no modification of the program needed.\\
&$-$ Difficult; not portable; limited to given data.\\
\hline
\end{tabular}
\end{table}

\subsubsection{Execution environment}

Many execution environments have interfaces with events from program
execution can be obtained. These interfaces include the Java Virtual
Machine Tool Interface (JVMTI~\cite{jvmti}) and
the LTTng interface for the Linux kernel~\cite{desnoyers2006lttng}.
Typically, the use of these mechanisms requires access to the data
structures of the execution environment itself. Monitoring code is
therefore not written at the level of the ``guest'' language (such
as Java in the case of JVMTI), but at the level of the ``host''
language, the language in which the virtual machine is written in.
This makes such monitoring approaches harder to use and less portable.
Finally, execution events may even be generated by custom hardware.

This technique has the advantage that it works in an unmodified execution
environment, and typically offers the least amount of overhead of all
approaches. However, even though modern environments offer a large range
of data through these tool interfaces, they are still restricted to
a predefined set of data.

Therefore, the option to use a special or modified execution environment
is sometimes used. A modified execution environment may change parts of
the kernel~\cite{artho2015using} or use a specific virtual machine that generates
a wider range of events~\cite{visser2003model}, or a debugger that can inspect
(and even modify) data at a finer level of detail than otherwise
possible~\cite{jakse2017interactive}.
Unlike standard environments, performance and stability of runtime
monitoring in these special environments depend on the modifications
needed to obtain the data.

Finally, a non-standard execution environment may even involve
special hardware access ports designed for monitoring~\cite{jtag}.
In this case, there is often no overhead, and at the software layer,
the execution environment is indistinguishable from a standard
environment.

Table~\ref{tab:instr} gives an overview of the different types of
approaches, and their pros and cons.
In~\cite{BartocciFFR18}, we further explain these concepts in two dedicated sections for hardware and software instrumentation.

%
%
\section{Interplay between Runtime Verification and other Verification Techniques}

\subsection{Static Techniques and Runtime Verification}
%
As opposed to runtime verification, static verification techniques are used to analyse and prove properties of all possible executions of programs. Two prominent families of static verification techniques are model checking and deductive verification. In the following, we will discuss both of them, together with their respective relation to runtime verification.

\subsubsection{Model Checking and Runtime Verification}
A popular verification technique besides runtime verification is \emph{model checking}~\cite{DBLP:books/daglib/0007403}.
While runtime verification checks a single execution of the underlying system, model checking considers all possible executions of the underlying system.
As such, both techniques share a common goal, i.e., the verification of an underlying system, but they have different features.
In general, the two techniques may be combined.
Let us briefly discuss the methodological combinations of the two techniques as well as their formal relationship.

At a first sight, model checking seems to provide stronger correctness guarantees than runtime verification as model checking considers all possible executions of a system.
However, model checking still greatly suffers from the so-called state-space explosion problem limiting its application only to parts of or abstractions of an underlying system.
As such, it makes perfectly sense to combine both verification techniques.
In \cite{DBLP:journals/jlp/LeuckerS09}, several combinations are discussed, which we briefly summarize here. 
\begin{itemize}
\item The verification result obtained by model checking is often referring to a model of the real system under analysis but not to the actual byte code, CPU etc.
\item However, the implementation might behave slightly different than predicted by the model. Runtime verification may then be used to easily check the actual execution of the system, to make sure that the implementation really meets its correctness properties.
	Thus, runtime verification may act as a partner to model checking.
\item
Often, some information is available only at runtime or is conveniently checked at runtime. For example, whenever library code with no accompanying source code is part of the system to build, only a vague description of the behavior of the code might be available. In such cases, runtime verification is an alternative to model checking.
\item The behavior of an application may depend heavily on the environment of the target system, for which certain assumptions have been taken at model checking time. Runtime verification allows to check such assumptions at runtime and may raise an alarm if one of the assumptions does not hold. Here, runtime verification completes the overall verification. 
\item In the case of systems where security is important or in the case of safety-critical systems, it is useful also to monitor behavior or properties that have been statically proved or tested, mainly to have a double check that everything goes well: Here again, runtime verification acts as a partner of model checking.
\end{itemize}
However, besides the methodological combination of runtime verification and model checking, there is a formal combination possible, as described in \cite{DBLP:conf/rv/Leucker12}. Let us recall the main idea here while we refer to the previously mentioned reference for formal details. 

Especially in online runtime verification, an execution of the underlying system is checked against a correctness property letter by letter. To reuse for example linear-time temporal logic (LTL) defined over infinite traces \cite{DBLP:conf/focs/Pnueli77}, the verdict for the finite execution is obtained by considering all possible infinite continuations of the execution. If with all such extensions the LTL formula yield either only \emph{true} or only \emph{false}, the corresponding verdict is that for the finite execution seen so far. Otherwise, the verdict is \emph{don't know?} to signal that the current excution does not allow a precise verdict anymore. This approach yields the so-called \emph{anticipatory} semantics introduced in \cite{DBLP:journals/tosem/BauerLS11}. 

In \cite{DBLP:conf/rv/Leucker12}, the idea was pursued that rather considering all possible infinite extensions of the current execution, only those are taken into account that yield runs of the overall system. Of course, a system model has to be at hand to understand which extension are possible at all. However, let us consider this idea for the empty execution, i.e., when the system has not even started: Then, we have to check whether all runs of the underlying system either satisfy or falsify the property at hand. The first question is actually the model-checking problem, and of course, if we have answered the model checking problem before we even start the execution of the system, the need for runtime verification vanishes. Now, assume that we consider an over-approximation of the underlying system, that is, a system having more runs/behavior, but smaller/less states. We still implicitly solve the model checking problem, indeed after every partial execution, yet for an abstract system. Thus, if the answer is that there is no violation possible any more (\emph{true}), one can stop monitoring. If, however, the model checking answer is \emph{false}, there might be a bug in the system, which may be found by further monitoring. 
Hence, by considering an over-approximation of the underlying system, we get best out of both worlds: we combine efficient runtime verification techniques with taking part of the system into account, hereby sharpening the verdict obtained by runtime verification. By tuning the abstraction, we can adjust whether the focus is on model checking or rather on runtime verification. In other words, by controlling the abstraction, we are able to slide between model checking and runtime verification. See \cite{DBLP:conf/rv/Leucker12} for further details. 
%
%
%

\subsubsection{Deductive Verification and Runtime Verification}
Deductive verification techniques are used to verify data-oriented, functional properties of code units (such as methods/procedures or classes), specified often in languages tailored to the programming language at hand (like, for instance, the Java Modeling Language ---JML).
Verification is typically done by reasoning directly about the source code, using a program logic like for instance Hoare logic or dynamic logic. Deductive program verification has been around for about 40 years, however, a number of developments during the last decade brought dramatic changes to how deductive verification is being perceived and used.
Among the state of the art efforts is the KeY tool for Java source code verification~\cite{lncs/10001}.
Deductive Verification has been extensively used to verify properties focusing on the system’s data at specific points of the execution (like method entries and exits).
Runtime verification, on the other hand, has been extensively used to verify trace properties with reasonable overheads (e.g., automata based or temporal). As both approaches work on the concrete system level, without abstraction, there is great potential for combining them.
Ideally, (sub)properties which are a bottleneck for static verification shall be addressed by runtime verification, whereas properties which require high overhead for runtime checking shall be addressed by static verification.
For that, however, we need specification languages allowing the expression of combined data- and control-flow properties in such a manner that they can be effectively decomposed for the application of different verification techniques.
This has been exemplified in the StaRVOOrS approach~\cite{AhrendtCPS17}, which provides a specification language combining data- and control-oriented aspects, and combines a deductive and a runtime verification tool to optimise runtime verification by (partial) results from (static) deductive verification.

%
%
\subsection{Model Learning and Runtime Verification}
%
Many verification techniques rely on the presence of a formal specification or a model of the system under verification.
In many cases, such a model is not available. Model learning is an approach to automatically infer the model, by observing traces of the system.
This can either be done passively, by purely observing the system, or actively, by steering the system execution into interesting areas.

Bertolino et al.~\cite{BertolinoCMS12} have suggested a combination of active learning and monitoring, where the produced system traces are continuously checked for conformance with the currently learned model. In case of non-conformance, the model will be updated to reflect the newly observed behaviour.
Isberner et al.~\cite{IsbernerHS14} proposed the TTT algorithm, which attempts to reduce the length of the observed counter example, to optimize the learning from long traces, as observed by runtime verification.

Both these approaches consider a continuous learning approach, where the learned model can be updated at any point of the execution.
Thus, if a bug is observed, it will be incorporated into the model.
Contrary to this, one could propose a two-step approach combining model learning and monitoring: in the first step the model is learned, using any of the existing algorithms.
In the second step, this learnt model is considered to be correct and used as a basis for the monitoring.
If, at any point during the runtime execution, an incorrect behaviour is encountered, it is reported as an anomalous behavior, rather than incorporated into the model.
The main goal of such an approach would be the detection of transient errors, i.e., errors that only occur sporadically or errors which only occur under certain conditions in the environment. 

Several techniques investigated how to learn different kinds of models that can enable the analysis of various classes of behaviors.
For instance, Mariani et al. investigated how to learn finite state models and likely program invariants to analyze executions in component-based systems \cite{MarianiPP11}.
Similarly, Pradel and Gross used learnt finite state models to analyze API misuses \cite{PradelG12}.
Recently, Grant et al. investigated how to infer and check assertions in distributed systems \cite{GrantCB18}.

\subsection{Testing and Runtime Verification}
%
To date, testing is by far the most commonly used technique to check software correctness. 
In essence, testing attempts to generate a number of test cases and checks that the outcome of each test case is as expected. While this technique is highly effective in uncovering bugs, it cannot guarantee that the tested system will behave correctly under all circumstances. 

Typically, testing is only employed during software development; meaning that software has no safety nets during runtime. Conversely, runtime verification techniques are typically used during runtime to provide extra correctness checks but little effort has been done to integrate it with the software development life cycle. For this reason little or no use of it is made in contemporary industry. 

At a closer look, the two techniques --- testing and runtime verification --- are intimately linked: runtime verification enables the checking of a system's behaviour during runtime by listening to system events, while testing is concerned with generating an adequate number of test cases whose behaviour is verified against an oracle. Thus, in summary one could loosely define runtime verification as event elicitation behaviour checking, and testing to be test case creation behaviour checking. 

Through this brief introduction of testing and runtime verification one can quickly note that behaviour checking is common to both. Given this overlap between two verification techniques, one is surprised to find that in the literature the two have not been well studied in each other's context, even though there were a few isolated attempts where RV and testing principles were jointly used~\cite{ArthoBGHKLPRSVW05,ArthoBS13,FalconeFMR06,FalconeFMR07}.

The paper~\cite{rvandtesting12}, outlines three ways in which this can be done: 
(i) one where testing can be used to support runtime verification in the creation of more reliable tools, 
(ii) another where the two techniques can be used together in a single tool such that the same set of properties can be tested and runtime verified, and 
(iii) a third approach where runtime verification can be used to support testing in gathering information regarding areas of the code which are observed executing at runtime.

%
\subsection{Runtime Assertion Checking and Runtime Verification}
%
%
Another area strongly connected to runtime verification (RV) is \emph{runtime assertion checking} (RAC).
Strictly speaking, RAC is a special case of RV.
Nonetheless, the characteristics are so different from most approaches to RV that RAC is sometimes seen as a field of its own.
In any case, we here take the approach which classifies RAC as an RV technique, whereas the other, mainstream approaches to RV are identified as \emph{runtime trace checking} (RTC).
We give an overview of this classification in Fig.~\ref{fig:camparetechniques}, also relating it to the major areas of static verification, model checking and deductive verification.
Note that we are deliberately simplifying the presentation.
Doing full justice to all verification approaches is not in scope of this section.
Rather we want to exhibit some basic characteristics.

\begin{table}[t]
\caption{Comparison of Verification Techniques (simplified)}
\centering
\begin{tabular}{|@{\hspace{1ex}}c@{\hspace{1ex}}|@{\hspace{1ex}}c@{\hspace{1ex}}|c|c|}
  \hline
  \textbf{Runtime Verification} & \textbf{Static Verification} & \textbf{Properties} & \textbf{Specifications}\\
  \hline
  \hline
  \specialcell{Runtime\\ Trace\\ Checking} & \specialcell{Model\\ Checking}
                                 & \specialcell{valid traces\\ \footnotesize{(+ some data)}}
                                 & \specialcell{temporal logics,\\ automata,\\ regular languages\\ (+ extensions)}\\
  \hline
  \specialcell{Runtime\\ Assertion\\ Checking} & \specialcell{Deductive\\ Verification}
                                         & \specialcell{valid data\\ in specific\\ code locations\\ \footnotesize{(+ some trace info)}}
                                         & \specialcell{first-order\\ assertion languages\\ (+ extensions)}\\
  \hline
\end{tabular}
\label{fig:camparetechniques}
\end{table}

RV mostly focuses on properties of execution \emph{traces} of some system, so most of RV fall under runtime trace checking (RTC).
These properties may be specified using different formalisms, of which temporal logics, automata, and regular languages are prominent examples.
In what concerns properties and specification approaches, RTC has therefore similarities with Model Checking. Most of this entire document is about RTC.

On the other hand, properties which are typically addressed by runtime assertion checking (RAC) focus on conditions on the \emph{data}, in specific code locations.
RAC comes with assertion languages, to formulate assertions to be `placed' at source code locations.
These assertions often use concepts from first-order logic to constrain which data is valid in the respective code locations.
RAC is therefore richer than RTC when it comes to properties of the data, but less expressive when it comes trace related properties.

A prominent example of a runtime assertion checker is OpenJML\footnote{\url{www.openjml.org}}, which supports the checking of JML~\cite{Huismanetal16} assertions while executing (instrumented) Java applications.
Other examples of RAC tools are SPARK~\cite{Barnes:2012:SPA:2436812}, and SPEC\#~\cite{Boogie04}, supporting variants of Ada and C\#, respectively.

One has to say that RAC techniques, because of the expressiveness of the used specification languages, are often too slow to be used for post-deployment RV, and are therefore better suited for the debugging phase.
This may be another reason why RAC is not always counted as an RV technique.
However, recent developments show that RAC can be very much optimized by combining static and runtime verification~\cite{AhrendtCPS17}.
%
%

%
%
\section{Challenges in Monitoring Quantitative and Statistical Data, beyond Property Violation}
%
Since runtime verification has traditionally borrowed techniques, and in particular specification languages, from static verification the specifications from which monitors are extracted are typically temporal logics and similar formalisms.
In turn, the outcome of a monitoring process is typically a Boolean verdict (or sometimes an enriched Boolean verdict to convey non-definite answers).
At the same time, decision problems on specifications (like equivalence vacuity, entailment, etc.) are decidable as these problems have been studied in static verification.
However, if one is willing to sacrifice the decidability of the these decision procedures, there is the possibility of designing “richer” logics.
One can perform runtime verification activities as long as there is a formal procedure to generate monitors from specifications, and an associated evaluation method for evaluating the generated monitors against an input trace.
Taking this abstract viewpoint on monitoring allows to consider languages that process rich data and generate richer outcomes. 

Consequently, we suggest here two directions in which runtime verification can be extended in terms of how rich the data that the monitors handle is.
The first direction is about the data the monitor processes and manipulates.
The second direction is about richer outcomes from the monitoring process.
%
\subsection{Richer Data for Monitors}
%
Two research areas related to runtime verification that study how to handle sequences of events with rich data, and that can produce rich data as outcome, are Complex Event Processing (CEP) and Data Stream Management Systems (DSMS); see~\cite{CugolaM12} for a modern survey.
Complex Event Processing considers distributed system that processes flows of events from different sources, finding correlations and producing derived events as a result. 
The main difference between CEP and RV (even the vision of RV with rich data that we advocate in this section) is that CEP does not provide a formal specification language with formal semantics, but instead an infrastructure to evaluate event processors. 
Similarly, DSMS borrow many techniques from Data Bases with the main difference that queries are supposed to process the flow of events without storing all events (even though many DSMS do create a storage with a time index).
Typically, DSMS process aggregations over flows of input events where the main concerns are efficiency in the evaluation process and statistical properties of the output (instead of logical correctness). 
One can potentially map the formal specifications that we use to generate monitors in RV to queries (for DSMS) and  to processors (in CEP), these areas do not provide formal translations nor a formal semantics of the execution platforms with guarantees on the order of arrival of events.

The book chapter “Monitoring Events that Carry Data”~\cite{HavelundRTZ18} 
\cite{DBLP:series/lncs/10457} reviews the research landscape on runtime verification techniques for traces that contain rich data, but we envision that richer and richer extensions will be investigated in years to come.
%
\subsection{Richer Outcomes from the Monitoring Process}
%
Another important aspect that has not been extensively studied in runtime verification is richer verdicts.
If one considers monitors as transformers from traces into useful “summaries” of information about the trace, there are many possibilities.
For example, in the area of signal temporal logics, there is the notion of “robustness”, which is a quantity that measures “how much” an observed trace matches a given specification (in specialized logics for the domain of signal processing like STL, MTL, etc.).
A value of 0 can indicate that  a full violation is detected, while values close to 0 (like 0.4) can indicate that a violation is close to occur.
This richer outcome allows to determine hot robustly a trace is satisfied or violated.
Potentially, the outcome of a monitor could be a pruned trace that allows further analysis to be performed, for example to determine the root of the error.
Finally, there are few works that consider the possibility of processing (statistically) incorrect or missing data.
One challenge for the near future is to explore different rich outcomes that the monitors can generate and the trade-offs involved in the evaluation of these monitors.

One potential direction to attack this challenge is Stream Runtime Verification (SRV), which was conceived to trigger richer verdicts beyond YES/NO answers.
SRV, pioneered by the tool Lola~\cite{DAngeloSSRFSMM05}, proposes to use streams to separate two concerns: the algorithms to perform evaluations of temporal properties on input traces, and the data collected during the evaluation.
The key insight is that many algorithms, for example to check LTL properties (and similar formalisms proposed to express monitors in the RV community) can be easily generalized to compute richer verdicts.
As a simple example, one can use the same algorithm that searches for a violating position in the past to compute the number of offending positions.
Similarly, one can compute the longest response time to a request, or the average response time (and not just the fact that all requests are answered). 
Modern extensions of stream runtime verification enrich the basic setting to parameterized properties~\cite{FaymonvilleFST16} so the input stream can be classified according to different objects.
Applications include network monitoring~\cite{FaymonvilleFST16}, security assurance of unmanned aerial vehicles~\cite{AdolfFFST17} and real-time SRV for the dynamic analysis of timed-event streams from low-level concurrent software~\cite{LeuckerSS0S18}.
All these results were motivated by challenging domains for which the ability of SRV to handle quantitative data and produce rich verdicts proved to be valuable.

Another potential direction to attack this challenge consists in using \emph{decentralised specifications}~\cite{El-HokayemF17}, pioneered by the tool THEMIS~\cite{El-HokayemF17a}.
While simple specifications can be expressed with traditional specification formalisms such as LTL and automata, accounting for hierarchies quickly becomes a problem.
Informally, a decentralized specification considers the system as a set of components, defines a set of monitors, additional atomic propositions that represent references to (the verdict of other) monitors, and attaches each monitor to a component.
Each monitor is a Moore automaton where the transition label is restricted to only atomic propositions related to the component on which the monitor is attached, and references to other monitors.
Decentralized specifications were successfully applied to the monitoring of smart homes~\cite{El-HokayemF18a} where they proved to allow a better scaling than with traditional specification formalisms.
%
%
%
%
\bibliographystyle{splncs}
\bibliography{references}
\end{document}